\documentclass[final,1p,times]{elsarticle}
\journal{E}
\usepackage{amsfonts}
\usepackage{tikz-cd}
\usepackage{extpfeil}
\usepackage{amssymb}
\usepackage{amsthm}
\usepackage{latexsym}
\usepackage{amsmath}
\usepackage{color}
\usepackage{graphicx}
\usepackage{indentfirst}
\usepackage{mathrsfs}
\usepackage{lipsum}
\allowdisplaybreaks

\newtheorem{theorem}{\color{black}\indent \textbf{Theorem}}[section]

\newtheorem{proposition}{\color{black}\indent Proposition}[section]
\newtheorem{definition}{\color{black}\indent Definition}[section]
\newtheorem{remark}{\color{black}\indent Remark}[section]

\newtheorem{example}{\color{black}\indent Example}[section]

\begin{document}
	
	\begin{frontmatter}

		\title{Higher-order and non-geometric symmetries and Noether's theorem in k-cosymplectic Hamiltonian field theory}

\author[a,b]{Manuel de León\corref{cor1}}
\author[c]{Xuefeng Zhao}
\address[a]{Instituto de Ciencias Matem\'aticas, Campus Cantoblanco, Consejo Superior de Investigaciones 
Científicas, C/ Nicolás Cabrera, 13–15, 28049, Madrid, Spain}
\address[b]{Real Academia Española de las Ciencias. C/ Valverde, 22, 28004, Madrid, Spain}
%\address[c]{Departamento de Matemática Aplicada, Universidad Politécnica de Madrid, Escuela de Edificación.  Av. Juan de Herrera 6, 28040 Madrid, Spain}
\address[c]{College of Mathematics, Jilin University, Changchun 130012, P. R. China}
\cortext[cor1]{Corresponding author}

		\begin{abstract}
		In this paper, we investigate higher-order symmetries and non-geometric symmetries within the framework of $k$-cosymplectic Hamiltonian field theory. We introduce the concept of generalized infinitesimal symmetry, which relaxes the standard symmetry condition, and then define two new classes of symmetries: higher-order general infinitesimal Cartan symmetries and good non-geometric symmetries. For each of these symmetry classes, we establish a Noether-type theorem that provides explicit conserved quantities for exact $k$-cosymplectic Hamiltonian systems. These results extend and generalize previous work on standard Cartan symmetries in the $k$-cosymplectic setting. Furthermore, we extend all of these results in a parallel fashion to Hamiltonian systems on $k$-symplectic manifolds, demonstrating the unifying nature of our approach. Concrete examples are provided to illustrate the existence and applicability of the proposed symmetries. Our framework offers a systematic geometric treatment of generalized symmetries and their associated conservation laws in classical field theories.
		\end{abstract}
		
	\end{frontmatter}
	\section{Introduction}
	In various branches of theoretical physics, the relationship between constants of motion and symmetries has long been a subject of considerable interest. In the contexts of classical Lagrangian mechanics and classical field theory, Noether's theorem \cite{Noether} undoubtedly stands as the most renowned contribution to this topic. Although this theorem has never truly faded from the scientific literature, it is fair to say that both the theorem itself and its various generalizations have experienced a resurgence of attention over the past decade. 
	
	The \(k\)-cosymplectic formalism, which extends the standard cosymplectic approach \cite{cantrijn,dLR89} (see also \cite{Cappelletti2,de4,Lucas}) to non-autonomous mechanics \cite{de1,de2}, offers a simple geometric setting for first-order classical field theories by incorporating base manifold coordinates into both the Lagrangian and Hamiltonian formalisms. Its mathematical foundation rests on the structure of \(k\)-cosymplectic manifolds \cite{Bazzoni,de1,de2,Gra,Rey2}. It should be noted, however, that this structure is distinct from the geometric structure of the recently developed \(q\)-cosymplectic manifolds \cite{Leok,Leok2}.
	
	Historically, this framework is rooted in the polysymplectic formalism introduced by Günther \cite{Gunther}, who proposed the notion of polysymplectic manifolds \cite{Blacker,Blacker2,Colombo,deLucas,Marrero2}. Independently, this concept was studied under the name of $k$-symplectic manifolds by A. Awane \cite{Awane1,Awane2,Awane3}, and also by de León and collaborators, in the latter case as a pure generalization of the canonical structures on tangent and cotangent bundles (see the book \cite{dLSV} for a complete account on that developments) (see also \cite{Cappelletti,deLeon2026geometric} for recent developments). The main issue on these structures is the existence of local coordinates of Darboux type \cite{de3, Awane1}. Over time, a variety of other polysymplectic approaches have also been developed for the geometric formulation of classical field theories \cite{Giachetta,Kanatchikov,McLean,Norris,Norris2}. Let us notice that the structures introduced by Norris are just $n$-symplectic structures on the frame bundle of an $n$-dimensional manifold. Correspondingly, the study of polycontact structures has also undergone vigorous development in recent years \cite{Finamore,Finamore2,Gaset2021,Leok3,deLucas2026,Zhao2}. %It is worth noting that a new \(q\)-contact structure has recently been developed, which constitutes a different geometric structure \cite{}.
	
	In the paper \cite{Roman}, the authors introduced the notion of Noether (or Cartan) symmetries,
	and stated Noether’s theorem for Hamiltonian and Lagrangian systems in $k$-symplectic field theories. Noether’s theorem associates conservation laws to the
	so-called Noether symmetries. However, as is known, in mechanics there are symmetries which are not of Noether type, but which also generate conserved quantities \cite{Lopez,Ranada,Ranada2}. Many attempts have been made to extend Noether’s theorem
	in order to include these symmetries and their conserved quantities for mechanical
	systems \cite{Echeverr,Sarlet}. Specially, \cite{Roman2} presented a particular family of non-Noether symmetries for $k$-symplectic Hamiltonian field theories.  
	
	In the present work, we extend the results of \cite{Roman2} to the framework of $k$-cosymplectic manifolds. Specifically, we introduce the concept of generalized infinitesimal symmetry and define the notion of a general infinitesimal Cartan symmetry of order $n$, thereby establishing a Noether theorem for general infinitesimal Cartan symmetries analogous to the results in \cite{Roman2}. Furthermore, inspired by the notion of non-geometric symmetries for classical Hamiltonian or Lagrangian systems as discussed in \cite{Roman3, Sarlet, Zhao}, we propose the concept of a non-geometric symmetry of order $n$ for $k$-cosymplectic Hamiltonian systems and derive a corresponding Noether theorem within this setting. In addition, we extend these results in a parallel fashion to Hamiltonian systems on $k$-symplectic manifolds.
	
	The paper is organized as follows. In Section 2, we review the fundamental concepts of 
	$k$-cosymplectic manifolds and the formulation of Hamiltonian systems within this geometric framework. Section 3 introduces the notion of symmetries for 
	$k$-cosymplectic Hamiltonian systems and analyzes their key properties. In Section 4, we define higher-order general infinitesimal Cartan symmetries and non-geometric symmetries, and establish the corresponding Noether theorems that yield conservation laws.  Section 5 extends these results in a parallel manner to Hamiltonian systems on 
	$k$-symplectic manifolds. Finally, Section~6 summarizes the main conclusions of this work, highlights the key novelties and differences from the existing literature, and discusses possible directions for future research.
	\section{$k$-Cosymplectic Hamiltonian Systems}
	We start by reviewing the core concepts of the theory of \(k\)-vector fields, as it provides the geometric foundation for describing systems of partial differential equations \cite{Rey}.
	\begin{definition}\label{def:kvecfield}
		A \(k\)-vector field on \(M\) is a smooth section \(\mathbf X: M \to \bigoplus^k TM\) of the vector bundle \(\text{pr}_M: \bigoplus^k TM \to M\). The set of all such sections is denoted by \(\mathfrak{X}^k(M)\).
	\end{definition}
	
	Any \(k\)-vector field \(\mathbf X \in \mathfrak{X}^k(M)\) can be identified with a \(k\)-tuple of ordinary vector fields \(X_1, \ldots, X_k \in \mathfrak{X}(M)\) via the relation \(X_\alpha = \text{pr}_M^\alpha \circ \mathbf X\) for each \(\alpha = 1, \ldots, k\). This identification allows us to write \(\mathbf X := (X_1, \ldots, X_k)\). Associated to \(\mathbf X\) there is a decomposable skew-symmetric contravariant tensor field \(X_1 \wedge \cdots \wedge X_k\), which constitutes a section of \(\bigwedge^k TM \to M\), as well as a generalised distribution \(D^X \subset TM\) generated by the vector fields \(X_1, \ldots, X_k\).
	
	\begin{definition}\label{def:prolongation}
		Given a smooth map \(\phi: U \subset \mathbb{R}^k \to M\), its first prolongation is the map \(\phi': U \subset \mathbb{R}^k \to \bigoplus^k TM\) given by
		\[
		\phi'(t)  = (\phi(t); \phi'_1(t),...,\phi'_k(t))= \left( \phi(t); T_t \phi\left(\frac{\partial}{\partial t^1}\bigg|_{t}\right), \ldots, T_t \phi\left(\frac{\partial}{\partial t^k}\bigg|_{t}\right) \right), 
		\]
        where $\phi'_\alpha(t)=T_t \phi\left(\frac{\partial}{\partial t^\alpha}\bigg|_{t}\right), \alpha=1,...,k,t = (t^1, \ldots, t^k) \in \mathbb{R}^k$
	\end{definition}
	
	Integral sections of \(k\)-vector fields are then introduced as follows.
	
	\begin{definition}\label{def:integralsection}
		Let \(\mathbf X = (X_1, \ldots, X_k) \in \mathfrak{X}^k(M)\) be a \(k\)-vector field. An integral section of \(\mathbf X\) is a map \(\phi: U \subset \mathbb{R}^k \to M\) satisfying \(\phi' =\mathbf X \circ \phi\), i.e.,
		\[
		T\phi\left(\frac{\partial}{\partial t^\alpha}\right) = X_\alpha \circ \phi \quad \text{for } \alpha = 1, \ldots, k.
		\]
		The \(k\)-vector field \(\mathbf X\) is said to be integrable if the underlying vector fields commute pairwise, i.e., \([X_\alpha, X_\beta] = 0\) for all \(1 \leq \alpha < \beta \leq k\).
	\end{definition}
	
	In local coordinates, write $X_\alpha = X^i_\alpha \frac{\partial}{\partial x^i}$ for $\alpha = 1, \ldots, k$. 
Then a map $\phi: U \subset \mathbb{R}^k \to M$ is an integral section of $\mathbf X$ precisely when its coordinate functions satisfy the system of first-order PDEs
\begin{align}\label{commut}
    \frac{\partial \phi^i}{\partial t^\alpha} = X^i_\alpha \circ \phi, \quad i = 1, \ldots, n, \quad \alpha = 1, \ldots, k. 
\end{align}

The compatibility condition for this overdetermined system is obtained by equating the mixed partial derivatives $\partial^2 \phi^i / \partial t^\alpha \partial t^\beta$ and $\partial^2 \phi^i / \partial t^\beta \partial t^\alpha$, which yields precisely
\[
[X_\alpha, X_\beta] = 0, \qquad 1 \leq \alpha < \beta \leq k. 
\]
\begin{remark}
    We emphasize that this is a rather strong integrability condition. It requires the vector fields $X_1,\ldots,X_k$ to commute pairwise, which is strictly stronger than the involutivity condition $[X_\alpha, X_\beta] \in \operatorname{span}\{X_1,\ldots,X_k\}$ required by Frobenius' theorem \cite{Lundell} for the existence of integral submanifolds. Indeed, for a multi-time PDE system of the form \eqref{commut}, pairwise commutativity is necessary and sufficient for the existence of local solutions for arbitrary initial data. 

This strong notion of integrability marks a fundamental distinction between $k$-symplectic/$k$-cosymplectic structures and multisymplectic structures. In the latter, one typically works with a distribution (e.g., the horizontal distribution of an Ehresmann connection \cite{de5,de6}) that is only required to be involutive in the Frobenius sense, or even merely locally generated by vector fields without being integrable. In contrast, the $k$-symplectic and $k$-cosymplectic settings impose the commutativity condition globally, reflecting the fact that the $k$ commuting vector fields provide a well-defined multi-time evolution.
\end{remark}

	Let us first recall the definition of $k$-symplectic manifold. As we know, the $k$-symplectic structures were introduced in \cite{Awane1,Gotay}.
	\begin{definition}[$k$-symplectic manifold]
		Let $M$ be a differentiable manifold of dimension $n+kn.$ A $k$-symplectic structure on $M$ is a family  $(\omega_0^A, V) (1\leq A\leq k)$, such that 
		$$\omega^A_0\in\Omega^2(M),\quad d\omega_0^A=0,\quad \cap_{A=1}^kker\omega_0^A=\{0\},$$
		and $ V$ is an integrable $nk$-dimensional tangent distribution on $M$ satisfying that
		$$\omega_0^A|_{ V\times V}=0,\quad \forall A.$$
		Then $(M,\omega_0^A, V)$ is called a $k$-symplectic manifold.
		
		The $k$-symplectic structure is exact if $\omega_0^A=d\theta_0^A$ for all $A.$
	\end{definition}
	%	Finally, we introduce a convenient notation. For \(\theta = \theta^\alpha \otimes e_\alpha \in \Omega^\ell(P, \mathbb{R}^k)\) and \(X = (X_1, \ldots, X_k) \in \mathfrak{X}^k(M)\), we define the contraction of \(\theta\) with \(X\) as
	%	\[
	%	\iota_X \theta := \langle \theta, X \rangle = \langle \theta^\alpha, X_\alpha \rangle := \iota_{X_\alpha} \theta^\alpha \in \Omega^{\ell-1}(M).
	%	\]
	\begin{theorem}[Darboux Theorem] (See \cite{de3}). Let $(\omega_0^A, V)$ be a $k$-symplectic structure on $M.$ For every point of $M,$ there exists a neighborhood $U$ and local coordinates $(q^i,p_i^A) (1\leq i\leq n,1\leq A\leq k)$ such that, on $U,$
		$$\omega_0^A=dq^i\wedge dp_i^A,\quad  V=\left<\frac{\partial}{\partial p_i^1},...,\frac{\partial}{\partial p_i^k}\right>_{i=1,...,n}.$$
		These are called Darboux or canonical coordinates of the $k$-symplectic manifold.	
	\end{theorem}
	Next, we recall the definition of $k$-cosymplectic manifold \cite{de1,Marrero,Munoz}.
	\begin{definition}[$k$-cosymplectic manifold]\label{Kco}
		Let $M$ be a differentiable manifold of dimension $k(n+1) + n$. A $k$-cosymplectic structure is a family $( \omega^A,\eta^A, \mathcal{V})$, $1 \leq A \leq k$, where $\eta^A \in \Omega^1(M)$, $\omega^A \in \omega^2(M)$, and $\mathcal{V}$ is an $nk$-dimensional distribution on $M$, such that
		\begin{enumerate}
			\item $\eta^1 \wedge \cdots \wedge \eta^k \neq 0$, $\eta^A|_{\mathcal{V}} = 0$, $\omega^A|_{\mathcal{V} \times \mathcal{V}} = 0$.
			\item $\left(\bigcap_{A=1}^k \ker \eta^A\right) \cap \left(\bigcap_{A=1}^k \ker \omega^A\right) = \{0\}$, $\dim\left(\bigcap_{A=1}^k \ker \omega^A\right) = k$.
			\item The forms $\eta^A$ and $\omega^A$ are closed, and $\mathcal{V}$ is integrable.
		\end{enumerate}
		Then $(M,  \omega^A,\eta^A, \mathcal{V})$ is called a $k$-cosymplectic manifold. It is termed an exact $k$-cosymplectic manifold if $\omega^A = d\theta^A$ for all $A$.
		
		If condition 2 does not hold, then we have a $k$-precosymplectic structure, and $(M,  \omega^A,\eta^A, \mathcal{V})$ is said to be a $k$-precosymplectic manifold.
	\end{definition}
	For every \( k \)-cosymplectic structure (\(  \omega^A,\eta^A, \mathcal{V} \)) on \( M \), there exists a family of \( k \) vector fields  
	\[\{ R_A \}_{1 \leq A \leq k},\]  
	which are called Reeb vector fields, characterized by the following conditions \cite{de1,Marrero}.  
	
	\[i(R_A) \eta^B = \delta^B_A, \quad i(R_A) \omega^B = 0; \quad 1 \leq A, B \leq k.\]
	\begin{theorem}(Darboux  Theorem) (See \label{Darboux} \cite{de1,Marrero}).
		If \( M \) is a \( k \)-cosymplectic manifold, then for every point of \( M \) there exists a local chart of coordinates \((t^A, q^i, p_i^A)\), \( 1 \leq A \leq k \), \( 1 \leq i \leq n \), such that
		
		\[
		\eta^A = dt^A, \quad \omega^A = dq^i \wedge dp_i^A, \quad \mathcal{V} = \left\langle \frac{\partial}{\partial p_i^1}, \dots, \frac{\partial}{\partial p_i^k} \right\rangle_{i=1,\dots,n}, \quad R_A = \frac{\partial}{\partial t^A}.
		\]
		
		These are called Darboux or canonical coordinates of the \( k \)-cosymplectic manifold.
	\end{theorem}
	Along this paper, we are interested in a kind of \( k \)-cosymplectic manifolds: those which are of the form \( \mathbb{R}^k \times M \), where \((M, \omega_0^A, V)\) is a generic \( k \)-symplectic manifold. Then, denoting by
	\[
	\pi_{\mathbb{R}^k} : \mathbb{R}^k \times M \to \mathbb{R}^k, \quad \pi_M : \mathbb{R}^k \times M \to M
	\]
	the canonical projections, we have the differential forms
	\[
	\eta^A = \pi_{\mathbb{R}^k}^* dt^A, \quad \omega^A = \pi_M^* \omega_0^A,
	\]
	and the distribution \( V \) in \( M \) defines a distribution \( \mathcal{V} \) in \(  \mathbb{R}^k \times M \) in a natural way. All the conditions given in definition \ref{Kco} are verified, and hence \( \mathbb{R}^k \times M \) is a \( k \)-cosymplectic manifold. From the Darboux Theorem \ref{Darboux} we have local coordinates \((t^A, q^i, p_i^A)\) in \( \mathbb{R}^k \times M \).
	
	%Observe that the standard model is a particular class of these kinds of \( k \)-cosymplectic manifolds, where \( M = (T_k^1)^* Q \).
	%	\begin{definition}
		%	These kinds of \( k \)-cosymplectic manifolds will be called almost-standard \( k \)-cosymplectic manifolds.	
		%	\end{definition}
	Consider an \( k \)-cosymplectic manifold \((\mathbb{R}^k \times M,  \omega^A,\eta^A, \mathcal{V})\), and let \( H \in C^\infty(\mathbb{R}^k \times M) \) be a Hamiltonian function. The couple \((\mathbb{R}^k \times M, H)\) is called a \( k \)-cosymplectic Hamiltonian system.
	
	We denote by \( X_H^k(\mathbb{R}^k \times M) \) the set of (local) \( k \)-vector fields \( X = (X_1, \ldots, X_k) \) on \( \mathbb{R}^k \times M \) which are solutions to the equations
	\begin{align}\label{Hamilton}
		\eta^A(X_B) = \delta_A^B, \quad \sum_{A=1}^k i(X_A)\omega^A = dH - \sum_{A=1}^k R_A(H)\eta^A.	
	\end{align}
	
	Since \( R_A = \partial/\partial t^A \) and \( \eta^A = dt^A \), then we can write locally the above equations as follows
	\[
	dt^A(X_B) = \delta_A^B, \quad \sum_{A=1}^k i(X_A)\omega^A = dH - \sum_{A=1}^k \frac{\partial H}{\partial t^A} dt^A.
	\]
	Furthermore, for a section \( \psi: I \subset \mathbb{R}^k \to \mathbb{R}^k \times M \) of the projection \( \pi_{\mathbb{R}^k} \), the Hamilton-de Donder-Weyl equations for this system are
	\begin{align}\label{HDW}
		\sum_{A=1}^k i_{ \left( \psi_*(t) \left( \frac{\partial}{\partial t^A} \Big|_t \right) \right) }(\omega^A \circ \psi) = \left[ dH - \sum_{A=1}^k R_A(H) \eta^A \right] \circ \psi ,
	\end{align}  	 
	
	\begin{remark}
		It should be noticed that, in general, equations \eqref{Hamilton} do not have a single solution. In fact, if $(\mathbb R^k\times M, \omega^A,\eta^A,\mathcal V)$ is a $k$-cosymplectic manifold we can define the vector bundle morphism,
		\begin{align*}
			\omega^\sharp:\;\quad T_k^1(\mathbb R^k\times M)&\rightarrow T^*(\mathbb R^k\times M)	\\
			(X_1,...,X_k)&\rightarrow\sum_{A=1}^ki_{X_A}\omega^A
		\end{align*}
		and, denoting by $\mathbb M_k(\mathbb R)$ the space of matrices of order $k$ whose entries are real numbers, the vector bundle morphism
		\begin{align*}
			\eta^\sharp:\;\quad T_k^1(\mathbb R^k\times M)&\rightarrow (\mathbb R^k\times M)\times\mathbb M_k(\mathbb R)	\\
			(X_1,...,X_k)&\rightarrow(pr_{\mathbb R^k\times M}(X_1,...,X_k),\eta^A(X_B)).
		\end{align*}
		We denote by the same symbols $\omega^\sharp,\eta^\sharp$ their natural extensions to vector fields and forms.
	\end{remark}
	
	%	Now, let $H:\mathbb R^k\times M\rightarrow\mathbb R$ be a real $C^\infty$-function on $\mathbb R^k\times M.$ Then, as in the case of an almost standard $k$-cosymplectic manifold, we can consider the set $\mathfrak{X}_H^k(\mathcal M)$ of the (local) $k$-vector fields $\mathbf X=(X_1,...,X_k)$ on $\mathcal M$ which are solutions to the equations
	%	\begin{align}\label{H2}
		%		\eta^A(X_B)=\delta^A_B,\quad\sum_{A=1}^ki_{X_A}\omega^A=dH-\sum_{A=1}^kR_A(H)\eta^A.
		%	\end{align}
	In \cite{Marrero}, the authors proved the following result:
	\begin{proposition}[\cite{Marrero}]\label{PP}
		The solutions to equation \eqref{Hamilton} are the sections of an affine bundle of rank $(k-1)(kn+n)$ which is modeled on the vector sub-bundle ker$\omega^\sharp\cap$ker$\eta^\sharp$ of $T^1_k (\mathbb R^k\times M).$
	\end{proposition}
\begin{remark}\label{R2.3}
	As noted in \cite{Marrero}, Proposition \ref{PP} implies the following. If $(X_1,\dots,X_k)$ is a particular solution of \eqref{Hamilton} and $\mathbf Z$ is a section of the vector bundle 
	$\ker \omega^\sharp \cap \ker \eta^\sharp \to \mathbb R^k \times M$, 
	then $(X_1,\dots,X_k) + \mathbf Z$ is again a solution of \eqref{Hamilton}. 
	Conversely, if $\mathbf X'$ and $\mathbf X$ are two solutions of \eqref{Hamilton}, then their difference 
	$\mathbf Z := \mathbf X' - \mathbf X$ is a section of the same vector bundle 
	$\ker \omega^\sharp \cap \ker \eta^\sharp \to \mathbb R^k \times M$.
\end{remark}
	
	\section{Symmetries For $k$-Cosymplectic Hamiltonian Systems}
	Let $(\mathbb R^k\times M,H)$ be a $k$-cosymplectic Hamiltonian system. First, following \cite{Marrero,Olver}, we introduce the next definition:
	\begin{definition}
		A conservation law for the Hamilton-de Donder-Weyl equations \eqref{HDW} is a map $\mathcal F=(F^1,...,F^k):\mathbb R^k\times M\rightarrow\mathbb R^k$ such that the divergence of
		$$\mathcal F\circ\psi=(F^1\circ\psi,...,F^k\circ\psi):U_0\subset\mathbb R^k\rightarrow\mathbb R^k$$
		is zero for every solution $\psi$ to the Hamilton-de Donder-Weyl equations \eqref{HDW}; that is for all $t\in U_0\subset\mathbb R^k,$
		\begin{align}
			0&=[Div(\mathcal F\circ\psi)](t)=\sum_{A=1}^k\frac{\partial(F^A\circ\psi)}{\partial t^A}|_t\nonumber\\
			&=\sum_{A=1}^k\psi_*(t)(\frac{\partial}{\partial t^A}|_t)(F^A).
		\end{align}
	\end{definition}
	In  \cite{Marrero}, the authors proved the following Proposition.
	\begin{proposition} \cite{Marrero}\label{PRL}
		The map $\mathcal F=(F^1,...,F^k):\mathbb R^k\times M\rightarrow \mathbb R^k$ defines a conservation law if and only if, for every integrable $k$-vector field $\mathbf X=(X_1,...,X_k)$ which is a solution to equation \eqref{Hamilton}, we have
		\begin{align}
			\sum_{A=1}^kL_{X_A}F^A=0.
		\end{align}
		
	\end{proposition}

	\begin{definition}\label{Infinite}
    \begin{enumerate}
        \item A \emph{symmetry} of the \(k\)-cosymplectic Hamiltonian system \((\mathbb{R}^k\times M,H)\) is a diffeomorphism
        \[
            \Phi \colon \mathbb{R}^k\times M \longrightarrow \mathbb{R}^k\times M
        \]
        satisfying the following conditions:
        \begin{enumerate}
            \item[(a)] it is a fiber-preserving map for the trivial bundle
            $\pi_{\mathbb{R}^k} \colon \mathbb{R}^k\times M \longrightarrow \mathbb{R}^k;$
            that is, \(\Phi\) induces a diffeomorphism \(\phi \colon \mathbb{R}^k \to \mathbb{R}^k\) such that
            \[
                \pi_{\mathbb{R}^k}\circ\Phi = \phi\circ\pi_{\mathbb{R}^k}.
            \]
            Equivalently, the following diagram commutes:
            \[
                \begin{tikzcd}
                    \mathbb{R}^k \times M \arrow[r, "\Phi"] \arrow[d, "\pi_{\mathbb{R}^k}"'] 
                        & \mathbb{R}^k \times M \arrow[d, "\pi_{\mathbb{R}^k}"] \\
                    \mathbb{R}^k \arrow[r, "\phi"'] 
                        & \mathbb{R}^k
                \end{tikzcd}
            \]

            \item[(b)] For every section \(\psi\) solution to the Hamilton--de Donder--Weyl equations \eqref{HDW}, the section
                $\Phi\circ\psi\circ\phi^{-1}$
            is also a solution to these equations.
        \end{enumerate}

        \item An \emph{infinitesimal symmetry} of the \(k\)-cosymplectic Hamiltonian system \((\mathbb{R}^k\times M,H)\) is a vector field
            $Y \in \mathfrak{X}(\mathbb{R}^k\times M)$
        whose flow consists of symmetries.
    \end{enumerate}
\end{definition}
	The following proposition gives a characterization of symmetries in terms of $k$-vector fields.
	\begin{proposition}\label{Pro5}
		Let $(\mathbb R^k\times M,H)$ be a $k$-cosymplectic Hamiltonian system and $\Phi:\mathbb R^k\times M\rightarrow \mathbb R^k\times M$ a fiber preserving diffeomorphism for the trivial bundle $\pi_{\mathbb R^k}:\mathbb R^k\times M\rightarrow \mathbb R^k.$
		
		1. For every integrable $k$-vector field $\mathbf X=(X_1,...,X_k)$ and for every integral section $\psi$ of $\mathbf X,$ the section $\Phi\circ\psi\circ\phi^{-1}$ is an integral section of the $k$-vector field $\Phi_*\mathbf X=(\Phi_*X_1,...,\Phi_*X_k),$ and hence $\Phi_*\mathbf X$ is integrable.
		
		2. $\Phi$ is a symmetry if and only if, for every integrable $k$-vector field $\mathbf X=(X_1,...,X_k)\in\mathfrak{X}_H^k(\mathbb R^k\times M),$ then $\Phi_*\mathbf X=(\Phi_*X_1,...,\Phi_*X_k)\in \mathfrak{X}_H^k(\mathbb R^k\times M).$
	\end{proposition}
	\begin{proof}
		See \cite{Marrero}.
	\end{proof}
	As a consequence of this, if $\Phi$ is a symmetry and $\mathbf X$ is an integrable $k$-vector field in $\mathfrak{X}_H^k(\mathbb R^k\times M),$ we have that $\Phi_*\mathbf X-\mathbf X\in$ker$\omega^\sharp\cap$ker$\eta^\sharp.$
	\begin{proposition}\label{Ker}
		Let $(\mathbb R^k\times M,H)$ be a $k$-cosymplectic Hamiltonian system. If $Y\in\mathfrak{X}(\mathbb R^k\times M)$ is an infinitesimal symmetry, then for every integrable $k$-vector field $\mathbf X=(X_1,...,X_k)\in \mathfrak{X}_H^k(\mathbb R^k\times M)$ we have that
		
		i)  $[Y,\mathbf X]\in$ker$\omega^\sharp\cap$ker$\eta^\sharp$; 
		 
		ii) For any $\mathbf W\in$ker$\omega^\sharp\cap$ker$\eta^\sharp$, then 
		$L_Y^n\mathbf W=(L_Y^nW_1,...,L_Y^nW_k)\in$ker$\omega^\sharp\cap$ker$\eta^\sharp,\forall n\geq1,$.
	\end{proposition}
	\begin{proof}
		Denote by $F_t$ the local $1$-parameter groups of diffeomorphisms generated by $Y.$ As $Y$ is an infinitesimal symmetry, as a consequence of Proposition \ref{Pro5}, we have $F_{t*}\mathbf X-\mathbf X\in $ker$\omega^\sharp\cap$ker$\eta^\sharp.$ Then, taking a local basis of sections $\{\mathbf Z^1,...,\mathbf Z^r\}=\{(Z_1^1,...,Z_k^1),...,(Z^r_1,...,Z_k^r)\}$ of the vector bundle ker$\omega^\sharp\cap$ker$\eta^\sharp\rightarrow \mathbb R^k\times M,$ we have that $F_{t*}\mathbf X-\mathbf X=g_\alpha\mathbf Z^\alpha,\alpha=1,...,r,$ with $g_\alpha:\mathbb R\times(\mathbb R^k\times M)\rightarrow \mathbb R$ (they are functions that depend on $t$); that is
		$$F_{t*}\mathbf X-\mathbf X=(F_{t*}X_1-X_1,...,F_{t*}X_k-X_k)=(g_\alpha Z_1^\alpha,...,g_\alpha Z_k^\alpha)=g_\alpha\mathbf Z^\alpha.$$
		Therefore
		\begin{align*}
			[Y,\mathbf X]&=L_Y\mathbf X=(L_YX_1,...,L_YX_k)=\left(lim_{t\rightarrow 0}\frac{F_{t*}X_1-X_1}{t},...,lim_{t\rightarrow 0}\frac{F_{t*}X_k-X_k}{t}\right)\\
			&=\left(lim_{t\rightarrow 0}\frac{g_\alpha}{t}Z_1^\alpha,...,lim_{t\rightarrow 0}\frac{g_\alpha}{t}Z_k^\alpha\right)=(f_\alpha Z_1^\alpha,...,f_\alpha Z_k^\alpha)=f_\alpha\mathbf Z^\alpha\in ker\omega^\sharp\cap ker\eta^\sharp,
		\end{align*}
		where $f_\alpha:\mathbb R^k\times M\rightarrow\mathbb R.$ 
		
		%Moreover, if    there exist integrable $k$-vector fields $\mathbf W,\mathbf W'\in\mathfrak{X}_H^k(\mathbb R^k\times M)$ such that $\mathbf W'-\mathbf W=\mathbf X,$  we know that $[Y,\mathbf X]=[Y,\mathbf W']-[Y,\mathbf W]\in$ker$\omega^\sharp\cap$ker$\eta^\sharp,$ since $[Y,\mathbf W'],[Y,\mathbf W]\in$ker$\omega^\sharp\cap$ker$\eta^\sharp$ by discussion above. Hence, we obtain that 
        %$$L_Y^2\mathbf X=L_Y[Y,\mathbf X]=[Y,f_\alpha \mathbf Z^\alpha].$$
        %Now, we prove that $[Y,f_\alpha \mathbf Z^\alpha]\in ker\omega^\sharp\cap ker\eta^\sharp.$ 
        
        For any $\mathbf W\in$ker$\omega^\sharp\cap$ker$\eta^\sharp$,
        we choose any $\mathbf V\in\mathfrak{X}_H^k(\mathbb R^k\times M),$ by Remark \ref{R2.3}, we know that $\tilde{\mathbf V}:=\mathbf V+\mathbf W\in \mathfrak{X}_H^k(\mathbb R^k\times M)$. Hence, by using i), 
        \begin{align*}
            [Y,\mathbf W]&=[Y,\tilde{\mathbf V}-\mathbf V]=[Y,\tilde{\mathbf V}]-[Y,\mathbf V]\in ker\omega^\sharp\cap ker\eta^\sharp.
        \end{align*}
		Similarly, we can get  $L_Y^n\mathbf W=(L_Y^nW_1,...,L_Y^nW_k)\in$ker$\omega^\sharp\cap$ker$\eta^\sharp,\forall n\geq1.$
	\end{proof}
	\begin{remark}
		It is easy to see that the converse of this proposition is not necessarily true, that is, i) and ii) may not necessarily imply that the  flows of $Y$ are symmetries.
	\end{remark}
	\section{Higher-order Cartan and Non-geometric symmetries. Noether's theorems}
	In \cite{Marrero}, it is shown that on a $k$-cosymplectic manifold, conservation laws can be associated with certain symmetries, known as infinitesimal Cartan or Noether symmetries. These are vector fields $Y\in\mathfrak{X}(\mathbb R^k\times M)$ satisfying: (a) $L_Y\omega^A=0$, (b) $i_Y\eta^A=0$, and (c) $L_YH=0$.
	The concept of higher-order infinitesimal Cartan symmetries for $k$-symplectic manifolds was introduced in \cite{Roman2}. In what follows, we extend this idea to the $k$-cosymplectic setting, aiming to construct a similar framework that yields conservation laws for $k$-cosymplectic Hamiltonian systems.

   According to Definition \ref{Infinite}, we have the notion of an infinitesimal symmetry for a \(k\)-cosymplectic Hamiltonian system. By Proposition \ref{Ker}, we know that for an arbitrary infinitesimal symmetry \(Y\) and the \(k\)-cosymplectic Hamiltonian vector field \(\mathbf{X}\), we can conclude that
\begin{align}\label{Infinite}
    [Y,\mathbf{X}] \in \ker\omega^\sharp \cap \ker\eta^\sharp .
\end{align}
Conversely, however, the flow of a vector field satisfying \eqref{Infinite} is not necessarily a symmetry. Hence, the set of vector fields satisfying \eqref{Infinite} is larger than the set of infinitesimal symmetry vector fields. Since it is precisely this more general class of vector fields that we shall consider in what follows, we introduce the notion of a generalized infinitesimal symmetry.
	\begin{definition}\label{De4}
		Let $(\mathbb R^k\times M,\omega^A,\eta^A,H)$ be a $k$-cosymplectic Hamiltonian system. A vector field $Y\in\mathfrak{X}(\mathbb R^k\times M)$ is said to be a generalized infinitesimal  symmetry for the $k$-cosymplectic Hamiltonian system if $L_Y\mathfrak{X}_H^k(\mathbb R^k\times M)\subset $ ker$\omega^\sharp\cap$ker$\eta^\sharp$, i.e.,
      $$[Y,\mathbf X]\in \ker\omega^\sharp\cap\ker\eta^\sharp,\quad \forall \mathbf X\in \mathfrak{X}_H^k(\mathbb R^k\times M).$$
	\end{definition}
	\begin{remark}\label{Re4}
		According to this definition, and arguing as in the proof of Proposition \eqref{Ker}~ii), we immediately obtain that, for every \(\mathbf{Z} \in \ker\omega^\sharp \cap \ker\eta^\sharp\),
\[
    L_Y^n \mathbf{Z} \in \ker\omega^\sharp \cap \ker\eta^\sharp
\]
for any \(n \ge 1\).
	\end{remark}
	\begin{definition}\label{DCar}
		Let $(\mathbb R^k\times M,\omega^A,\eta^A,H)$ be a $k$-cosymplectic Hamiltonian system. A vector field $Y\in\mathfrak{X}(\mathbb R^k\times M)$ is said to be a general infinitesimal Cartan or Noether symmetry of order $n$ if:
		
		1. $Y$ is a generalized infinitesimal symmetry;
		
		2. $i_Y\eta^A=0;$
		
		3. \( L^{n}_Y(\omega^1,...,\omega^k):=(\overbrace{L_Y \cdots L_Y}^{n}\omega^1,...,\overbrace{L_Y \cdots L_Y}^{n}\omega^k)= \vec0 \), but \( L^m_Y (\omega^1,...,\omega^k) \neq \vec 0 \), for \( m < n \);
		
		4. $ L_Y^nH=0.$ 
	\end{definition} 
	\begin{remark}
		For $n=1$ and $Y$ is an infinitesimal symmetry, we recover the definition of an infinitesimal Cartan (Noether) symmetry \cite{Marrero}. However, when $n=1$ and $Y$ only is a generalized infinitesimal symmetry,  $Y$ may not be  infinitesimal Cartan symmetries.
	\end{remark}
	
	\begin{remark}
		According to \cite{Roman2}, we know that for a $k$-symplectic Hamiltonian system, vector field $Y$ is said to be an infinitesimal Cartan or Noether symmetry of order $n$ if (a) $Y$ is an infinitesimal symmetry, (b) \( L^n_Y\omega^A:=\underbrace{L_Y \cdots L_Y}_{n}\omega^A= 0 \), but \( L^m_Y \omega^A \neq 0 \), for \( m < n \) and (c) $L_YH=0.$ In the case of $k$-cosymplectic Hamiltonian system, we give a additional condition $i_Y\eta^A=0$ and modify the conditions (a) and (c) to   looser conditions 1 and 4. In fact, in \cite{Roman2}, conditions $Y$ is an infinitesimal symmetry and $L_YH=0$ can also be generalized to conditions 1 and 4, which will be redefined in next section.
	\end{remark}
	
	\begin{proposition}\label{Pro4}
		Let $Y\in\mathfrak{X}(\mathbb R^k\times M)$ be an general infinitesimal $k$-cosymplectic Noether symmetry of order $n$ of an exact $k$-cosymplectic Hamiltonian system $(\mathbb R^k\times M,\omega^A=d\theta^A,\eta^A,H)$. Then, for every $p\in \mathbb R^k\times M,$ there exists an open neighborhood $U_p\ni p,$ such that:
		
		1. There exist functions $F^A\in C^\infty(U_p),$ unique up to additive constants, satisfying
		\begin{align}\label{LYom}
			L_Y^{n-1}(i_Y\omega^A)=dF^A\quad \text{on } U_p.
		\end{align}
		
		2. There exist functions $\zeta^A\in C^\infty(U_p)$ such that $L_Y^n\theta^A=d\zeta^A$ on $U_p,$ and consequently
		\begin{align}\label{zeta}
			\zeta^A=F^A+L_Y^{n-1}i_Y\theta^A \quad (\text{up to an additive constant on } U_p).
		\end{align}
		
	\end{proposition}
	\begin{proof}
		1. This follows from the Poincaré Lemma together with the condition
		\[
		0 = L_Y^n \omega^A = L_Y^{n-1} L_Y \omega^A = L_Y^{n-1} (d i_Y \omega^A) = d L_Y^{n-1} (i_Y \omega^A).
		\]
		
		2. We have
		\[
		d L_Y^n \theta^A = L_Y^n d\theta^A = L_Y^n \omega^A = 0,
		\]
		so the $L_Y^n \theta^A$ are closed forms. By the Poincaré Lemma, there exist $\zeta^A \in C^\infty(U_p)$ such that $L_Y^n \theta^A = d\zeta^A$ on $U_p$. Moreover, since \eqref{LYom} holds on $U_p$, we obtain
		\begin{align*}
			d\zeta^A &= L_Y^n \theta^A = L_Y^{n-1} L_Y \theta^A = L_Y^{n-1} (d i_Y \theta^A + i_Y d\theta^A) \\
			&= d L_Y^{n-1} i_Y \theta^A + L_Y^{n-1} (i_Y \omega^A) = d\bigl( L_Y^{n-1} i_Y \theta^A + F^A \bigr),
		\end{align*}
		which implies \eqref{zeta}.
	\end{proof}

	Now, we state that Noether's theorem can be generalized for these higher-order Cartan symmetries and $k$-cosymplectic Hamiltonian systems as follows:
	\begin{theorem}[Noether theorem for general infinitesimal Cartan symmetries]\label{NoG} If $Y\in\mathfrak{X}(\mathbb R^k\times M)$ is a general infinitesimal Cartan symmetry of order $n$ of an exact $k$-cosymplectic Hamiltonian system $(\mathbb R^k\times M,\omega^A=d\theta^A,\eta^A,H)$, then
		$$\mathcal F=(F^1,...,F^k)=(\zeta^1-L_Y^{n-1}i_Y\theta^1,...,\zeta^k-L_Y^{n-1}i_Y\theta^k)$$ 
		is a conserved quantity, that is, for every integrable $k$-vector field $\mathbf X=(X_1,...,X_k)\in \mathfrak{X}_H^k(\mathbb R^k\times M),$ we have that $\sum_{A=1}^kL_{X_A}F^A=0$ (on $U_p$).	
	\end{theorem}
	\begin{proof}
		If $\mathbf X=(X_1,...,X_k)\in \mathfrak{X}_H^k(\mathbb R^k\times M),$ taking \eqref{LYom} into account we have
		$$\sum_{A=1}^kL_{X_A}F^A=\sum_{A=1}^ki_{X_A}dF^A=\sum_{A=1}^ki_{X_A}L_Y^{n-1}(i_Y\omega^A).$$
		Then, if $n=2,$  we have 
		\begin{align*}
			\sum_{A=1}^kL_{X_A}F^A&=\sum_{A=1}^ki_{X_A}L_Y(i_Y\omega^A)=\sum_{A=1}^k(L_Yi_{X_A}-i_{[Y,X_A]})i_Y\omega^A\\
			&=\sum_{A=1}^k(-L_Yi_Yi_{X_A}+i_Yi_{[Y,X_A]})\omega^A\\
			&=-L_Yi_Y(dH-\sum_{A=1}^k(R_AH)\eta^A)\\
			&=-L^2_YH=0.
		\end{align*}
		since $Y$ is an infinitesimal Cartan symmetry of order $n,L_Y^2H=0$ and by Definition\ref{De4}, $[Y,\mathbf X]\in Ker\omega^\sharp\cap\eta^\sharp,$ and hence $\sum_{A=1}^ki_{[Y,X_A]}\omega^A=0.$ 
		
		If $n=3,$ by an analogous reasoning, we obtain
		\begin{align*}
			\sum_{A=1}^k L_{X_A} F^A &= \sum_{A=1}^k i_{X_A} L_Y L_Y (i_Y \omega^A) = \sum_{A=1}^k (L_Y i_{X_A} - i_{[Y,X_A]}) L_Y i_Y \omega^A \\
			&= \sum_{A=1}^k \left( L_Y i_{X_A} L_Y - i_{[Y,X_A]} L_Y \right) i_Y \omega^A \\
			&= \sum_{A=1}^k \left( L_Y^2 i_{X_A}  - 2L_Y i_{[Y,X_A]} + i_{[Y,[Y,X_A]]} \right) i_Y \omega^A \\
			&= \sum_{A=1}^k L_Y^{2} i_{X_A} i_Y \omega^A = -\sum_{A=1}^k L_Y^{2} i_Y i_{X_A} \omega^A \\
			&= -L_Y^{2} i_Y \left( dH - \sum_{A=1}^k (R_A H) \eta^A \right) \\
			&= -L_Y^3 H = 0.
		\end{align*}
		
		For $n>3,$ we arrive at the same result by repeating the above procedure $n-2$ times. In fact,
		\begin{align*}
			\sum_{A=1}^k L_{X_A} F^A &= \sum_{A=1}^k i_{X_A} L_Y L_Y^{n-2} (i_Y \omega^A) = \sum_{A=1}^k (L_Y i_{X_A} - i_{[Y,X_A]}) L_Y^{n-2} i_Y \omega^A \\
			&= \sum_{A=1}^k \left( L_Y i_{X_A} L_Y^{n-2} - i_{[Y,X_A]} L_Y^{n-2} \right) i_Y \omega^A \\
			&= \sum_{A=1}^k \left( L_Y^2 i_{X_A} L_Y^{n-3} - 2L_Y i_{[Y,X_A]} L_Y^{n-3} + i_{[Y,[Y,X_A]]} L_Y^{n-3} \right) i_Y \omega^A \\
			&= \sum_{A=1}^k \left( L_Y^3 i_{X_A} L_Y^{n-4} - L_Y^2 i_{[Y,X_A]} L_Y^{n-4} - 2L_Y^2 i_{[Y,X_A]} L_Y^{n-4} + 2L_Y i_{[Y,[Y,X_A]]} L_Y^{n-4} - i_{[Y,[Y,[Y,X_A]]]} L_Y^{n-4} \right) i_Y \omega^A \\
			&= \sum_{A=1}^k \left( L_Y^{n-1} i_{X_A} + c_1 L_Y^{n-2} i_{[Y,X_A]} + c_2 L_Y^{n-3} i_{L_Y^2 X_A} + \cdots + c_{n-1} i_{L_Y^{n-1} X_A} \right) i_Y \omega^A \\
			&= \sum_{A=1}^k L_Y^{n-1} i_{X_A} i_Y \omega^A = -\sum_{A=1}^k L_Y^{n-1} i_Y i_{X_A} \omega^A \\
			&= -L_Y^{n-1} i_Y \left( dH - \sum_{A=1}^k (R_A H) \eta^A \right) \\
			&= -L_Y^n H = 0,
		\end{align*}
		where $c_1,...,c_{n-1}$ are constants.
		Thus, taking into account Remark \ref{Re4}, we have proved that $\mathcal F=(F^1,...,F^k)$ is a conservation law.
	\end{proof}
	\begin{example}\label{Ex1}
		Consider a $2$-cosymplectic manifold defined as follows:
		\[
		M = \mathbb{R} \times \mathbb{R}^2 \setminus \{0\}, \quad 
		\eta^1 = dt^1,\quad 
		\eta^2 = dt^2,\quad 
		\omega^1 = p_1 \, dq \wedge dp_1,\quad 
		\omega^2 = dq \wedge dp_2,\quad 
		\mathcal{V} = \operatorname{span}\left\{ \frac{\partial}{\partial p_1}, \frac{\partial}{\partial p_2} \right\},
		\]
		where $\mathbb{R}^2 \times M$ has coordinates $(t^1, t^2, q, p_1, p_2)$. Let 
		\[
		\theta^1 = q \, d(p_1^2/2), \quad \theta^2 = q \, dp_2,
		\]
		it is clear that $d\theta^1 = \omega^1$ and $d\theta^2 = \omega^2$. We can compute that
		\[
		\ker \omega^\sharp \cap \ker \eta^\sharp = 
		\operatorname{span}\left\{ \left( f\frac{\partial}{\partial p_1} + g\frac{\partial}{\partial p_2}, \; h\frac{\partial}{\partial p_1} - p_1 f\frac{\partial}{\partial p_2} \right) \; \middle| \; \forall f,g,h \in C^\infty(\mathbb{R}^2 \times M) \right\}.
		\]
		
		Assume the Hamiltonian function $H = p_1 + t^1 + t^2$. Using equation \eqref{Hamilton}, we obtain
		\[
		(X_1, X_2) = \left( \frac{\partial}{\partial t^1} + \frac{1}{p_1} \frac{\partial}{\partial q}, \; \frac{\partial}{\partial t^2} + \frac{\partial}{\partial p_1} \right) \in \mathfrak{X}_H^2(\mathbb{R}^2 \times M).
		\]
		
		Let $Y = \frac{\partial}{\partial q}$. A straightforward computation shows that
		\[
		L_Y(X_1, X_2) = 0, \quad L_Y\left( \ker \omega^\sharp \cap \ker \eta^\sharp \right) \subset \ker \omega^\sharp \cap \ker \eta^\sharp,
		\]
		hence $Y$ is a generalized infinitesimal symmetry. Moreover, we can compute
		\[
		i_Y \eta^1 = i_Y \eta^2 = 0, \quad L_Y \omega^1 = L_Y \omega^2 = 0, \quad L_Y H = 0.
		\]
		Thus, $Y$ is a general infinitesimal Cartan symmetry of order $1$. Furthermore, we have
		\[
		L_Y \theta^1 = d\left(\frac{p_1^2}{2}\right), \quad L_Y \theta^2 =  dp_2  , \quad i_Y \theta^1 = 0, \quad i_Y \theta^2 = 0.
		\]
		We therefore choose $F^1 = \zeta^1 - i_Y \theta^1 = \frac{p_1^2}{2}$, $F^2 =\zeta^2-i_Y\theta^2= p_2$. It is evident that
		\[
		L_{X_1} F^1 + L_{X_2} F^2 = 0.
		\]
		Hence, $(F^1,F^2)=(\frac{p_1^2}{2}, p_2)$ is a conserved quantity.
	\end{example}
	In \cite{Roman3,Sarlet,Zhao}, we can find the notion of non-geometric symmetries for classical Hamiltonian systems or Lagrangian systems. Specially, in \cite{Roman3}, the author considered the non-geometric symmetries of Hamiltonian systems $(M,\omega,H)$ on symplectic manifold $(M,\omega)$, which satisfy
	$$[Y,X_H]=0,\quad L_Y^n\omega=f_0\omega+f_1L_Y\omega+\cdots+f_{n-1}L_Y^{n-1}\omega,\;f_i\in C^\infty(M),i=0,...,n-1.$$
	In this paper, we  introduce non-geometric  symmetries for Hamiltonian systems on $k$-cosymplectic manifold.
	\begin{definition}\label{Dgeo}
	Let $(\mathbb{R}^k\times M,\omega^A=d\theta^A,\eta^A,H)$ be an exact $k$-cosymplectic Hamiltonian system. A vector field $Y\in\mathfrak{X}(\mathbb{R}^k\times M)$ is said to be a non-geometric symmetry of order $n$ if:
	
	\begin{enumerate}
		\item $Y$ is a generalized infinitesimal symmetry;
		
		\item $i_Y\eta^A=0$;
		
		\item $n$ is the first number satisfying
		\[
		L^{n}_Y\omega^A:=\overbrace{L_Y \cdots L_Y}^{n}\omega^A
		=c_0\omega^A+ c_1L_Y\omega^A+\cdots+c_{n-1}L_Y^{n-1}\omega^A,
		\]
		where $c_i\in \mathbb{R}$, $i=0,\dots,n-1$, and
		\[
		L_YH=0.
		\]
	\end{enumerate}
	
	Specially, if $c_0=0$, we say that $Y$ is a good non-geometric symmetry of order $n$.
\end{definition}
\begin{remark}\label{rem:CartanVsNonGeometric}
	We compare Definition~\ref{DCar} (general infinitesimal Cartan or Noether
	symmetry of order \(n\)) with Definition~\ref{Dgeo} (non-geometric symmetry
	of order \(n\)). Although both notions are formulated for Hamiltonian systems
	on \(\mathbb{R}^k\times M\) and both impose
	\begin{equation}\label{eq:common}
		i_Y\eta^A=0
	\end{equation}
	together with a condition on the Lie derivatives of the Hamiltonian function
	\(H\), they differ in two respects: in the behaviour required of the
	higher-order Lie derivatives of the presymplectic forms \(\omega^A\), and in
	the order of the Lie derivative imposed on \(H\).

	Recall that in Definition~\ref{DCar} one requires
	\begin{equation}\label{eq:CartanForm}
		L_Y^n(\omega^1,\dots,\omega^k)
		:=\bigl(L_Y^n\omega^1,\dots,L_Y^n\omega^k\bigr)=\vec 0,
		\qquad
		L_Y^m(\omega^1,\dots,\omega^k)\neq \vec 0
		\quad (m<n),
	\end{equation}
	and
	\begin{equation}\label{eq:CartanH}
		L_Y^nH=0.
	\end{equation}
	Thus a Cartan/Noether symmetry of order \(n\) is characterized by the
	\emph{simultaneous vanishing} of all \(n\)-th Lie derivatives of the forms
	\(\omega^A\), with \(n\) being the first order at which this vanishing
	occurs, and by the vanishing of the \(n\)-th Lie derivative of \(H\).

	In Definition~\ref{Dgeo}, on the other hand, one requires that \(n\) be the
	first number such that
	\begin{equation}\label{eq:geoForm}
		L_Y^n\omega^A
		=c_0\omega^A+c_1L_Y\omega^A+\cdots+c_{n-1}L_Y^{n-1}\omega^A,
		\qquad c_i\in\mathbb{R},
	\end{equation}
	and that
	\begin{equation}\label{eq:geoH}
		L_YH=0.
	\end{equation}
	That is, for \(H\) only the \emph{first-order} Lie derivative is required to
	vanish, not the \(n\)-th one. For the forms \(\omega^A\), one
	requires \(L_Y^n\omega^A\) to lie in the linear span of the lower-order Lie
	derivatives
	\(\omega^A,L_Y\omega^A,\dots,L_Y^{n-1}\omega^A\), rather than to vanish.
	When \(c_0=\cdots=c_{n-1}=0\), condition~\eqref{eq:geoForm} reduces to
	\(L_Y^n\omega^A=0\), and in this case \(Y\) is also a \emph{good}
	non-geometric symmetry of order \(n\). Hence, generally we get that \emph{Cartan/Noether \(\not\Rightarrow\) non-geometric symmetry} and \emph{Non-geometric \(\not\Rightarrow\) Cartan/Noether in general.} However, it can be seen that  general infinitesimal Cartan symmetries of order 1 are just the good non-geometric symmetries of order 1.
\end{remark}
	
	We give the following example to illustrate that such non-geometric symmetry does indeed exist.
	\begin{example}\label{Ex2}
		Consider a $2$-cosymplectic manifold defined as follows:
		\[
		M = \mathbb{R} \times \mathbb{R}^2\setminus\{0\}, \quad 
		\eta^1 = dt^1,\quad 
		\eta^2 = dt^2,\quad 
		\omega^1 = e^{cp_1} \, dq \wedge dp_1,\quad 
		\omega^2 = dq \wedge dp_2,\quad 
		\mathcal{V} = \operatorname{span}\left\{ \frac{\partial}{\partial p_1}, \frac{\partial}{\partial p_2} \right\},
		\]
		where $\mathbb{R}^2 \times M$ has coordinates $(t^1, t^2, q, p_1, p_2)$. Let 
		\[
		\theta^1 = q \, d(e^{cp_1}/c), \quad \theta^2 = q \, dp_2,
		\]
		it is clear that $d\theta^1 = \omega^1$ and $d\theta^2 = \omega^2$. We can compute that
		\[
		\ker \omega^\sharp \cap \ker \eta^\sharp = 
		\operatorname{span}\left\{ \left( f\frac{\partial}{\partial p_1} + g\frac{\partial}{\partial p_2}, \; h\frac{\partial}{\partial p_1} - e^{cp_1} f\frac{\partial}{\partial p_2} \right) \; \middle| \; \forall f,g,h \in C^\infty(\mathbb{R}^2 \times M) \right\}.
		\]
		
		Assume that the Hamiltonian function is given by \(H = \kappa(t_1, t_2)\), where \(\kappa\) is an arbitrary function of \(t_1\) and \(t_2\). Using equation \eqref{Hamilton}, we obtain
		\[
		(X_1, X_2) = \left( \frac{\partial}{\partial t^1}, \; \frac{\partial}{\partial t^2} - \frac{\partial}{\partial p_1} \right) \in \mathfrak{X}_H^2(\mathbb{R}^2 \times M).
		\]
		
		Let $Y = q\frac{\partial}{\partial q}$. A straightforward computation shows that
		\[
		L_Y(X_1, X_2) = \left(0,0\right), \quad L_Y\left( \ker \omega^\sharp \cap \ker \eta^\sharp \right) \subset \ker \omega^\sharp \cap \ker \eta^\sharp,
		\]
		hence $Y$ is a generalized infinitesimal symmetry. Moreover, we can compute
		\[
		i_Y \eta^1 = i_Y \eta^2 = 0, \quad L_Y^1 \omega^1=\omega^1, \quad L^1_Y \omega^2 = \omega^2, \quad L_Y^2 H = 0.
		\]
		Thus, $Y$ is a non-geometric symmetry of order $1$. 
	\end{example}
    \begin{remark}
        According to the definition of Hamiltonian systems on \(k\)-cosymplectic manifolds, we know that the Hamiltonian vector field corresponding to any function depending only on the time variables is, in essence, an element of \(\ker \omega^\sharp \cap \ker \eta^\sharp\) together with a vector field along the time direction. Therefore, the example given above can essentially be understood as a good non-geometric symmetry of elements in \(\ker \omega^\sharp \cap \ker \eta^\sharp\). Consequently, when we are interested in the first integrals of elements in \(\ker \omega^\sharp \cap \ker \eta^\sharp\), we can also make use of the good non-geometric symmetry defined in this paper, combined with the method described in the propositions below, to obtain these first integrals.
    \end{remark}

	Similar to the conclusion of Proposition \ref{Pro4}, for non-geometric symmetries of order $n$, we can obtain the conclusion of the following proposition.
	\begin{proposition}\label{Propo4.2}
		Let $Y\in\mathfrak{X}(\mathbb R^k\times M)$ be an non-geometric symmetry  of order  $n$ of an exact $k$-cosymplectic Hamiltonian system $(\mathbb R^k\times M,\omega^A=d\theta^A,\eta^A,H)$. Then, for every $p\in \mathbb R^k\times M,$ there exists an open neighborhood $U_p\ni p,$ such that:
		
		1. There exist functions $F^A\in C^\infty(U_p),$ unique up to additive constants, satisfying
		\begin{align}\label{LYomm}
			L_Y^{n-1}i_Y\omega^A-c_{n-1}L_Y^{n-2}i_Y\omega^A-\cdots-c_1i_Y\omega^A-c_0\theta^A=dF^A\quad \text{on } U_p.
		\end{align}
		
		2. There exist functions $\zeta^A\in C^\infty(U_p)$ such that $L_Y^n\theta^A=d\zeta^A$ on $U_p,$ and consequently
		\begin{align}\label{zetaa}
			\zeta^A=F^A+L_Y^{n-1}  i_Y \theta^A-c_{n-1}L_Y^{n-2} i_Y \theta^A-\cdots-c_1 i_Y \theta^A \quad (\text{up to an additive constant on } U_p).
		\end{align}
		
	\end{proposition}
	\begin{proof}
		1. Firstly, we can calculate that
		\begin{align*}
			L_Y^n \omega^A &= L_Y^{n-1} L_Y \omega^A = L_Y^{n-1} (d i_Y \omega^A) = d L_Y^{n-1} (i_Y \omega^A),\\
			L_Y^{n-1} \omega^A &= L_Y^{n-2} L_Y \omega^A = L_Y^{n-2} (d i_Y \omega^A) = d L_Y^{n-2} (i_Y \omega^A),\\
			&\vdots\\
			L_Y\omega^A&=di_Y\omega^A,
		\end{align*}
		thus,
		\begin{align*}
			0&=L_Y^n\omega^A-c_{n-1}L_Y^{n-1}\omega^A-\cdots-c_1L_Y\omega^A-c_0\omega^A\\
			&=d(L_Y^{n-1}i_Y\omega^A-c_{n-1}L_Y^{n-2}i_Y\omega^A-\cdots-c_1i_Y\omega^A-c_0\theta^A).
		\end{align*}
		By Poincaré Lemma, we can get the result 1.
		
		2. Secondly, we have
		\[
		d L_Y^i \theta^A = L_Y^i d\theta^A = L_Y^i \omega^A,\quad i=0,1,...,n,
		\]
		so $L_Y^n \theta^A-c_{n-1}L_Y^{n-1}\theta^A-\cdots-c_1L_Y\theta^A-c_0\theta^A$ are closed forms. By the Poincaré Lemma, there exist $\zeta^A \in C^\infty(U_p)$ such that $L_Y^n \theta^A-c_{n-1}L_Y^{n-1}\theta^A-\cdots-c_1L_Y\theta^A-c_0\theta^A = d\zeta^A$ on $U_p$. Moreover, since \eqref{LYomm} holds on $U_p$, we obtain
		\begin{align*}
			d\zeta^A &= L_Y^n \theta^A-c_{n-1}L_Y^{n-1}\theta^A-\cdots-c_1L_Y\theta^A-c_0\theta^A \\
			& = L_Y^{n-1} L_Y \theta^A -c_{n-1}L_Y^{n-2}L_Y\theta^A-\cdots-c_1L_Y\theta^A-c_0\theta^A \\
			&= L_Y^{n-1} (d i_Y \theta^A + i_Y d\theta^A)-c_{n-1}L_Y^{n-2}(d i_Y \theta^A + i_Y d\theta^A)-\cdots-c_1(d i_Y \theta^A + i_Y d\theta^A)-c_0\theta^A \\
			&= d(L_Y^{n-1}  i_Y \theta^A-c_{n-1}L_Y^{n-2} i_Y \theta^A-\cdots-c_1 i_Y \theta^A )\\
			&\quad+(L_Y^{n-1}i_Y\omega^A-c_{n-1}L_Y^{n-2}i_Y\omega^A-\cdots-c_1i_Y\omega^A-c_0\theta^A) \\
			&=   d(L_Y^{n-1}  i_Y \theta^A-c_{n-1}L_Y^{n-2} i_Y \theta^A-\cdots-c_1 i_Y \theta^A + F^A ),
		\end{align*}
		which implies \eqref{zetaa}.
	\end{proof}
	For good non-geometric symmetries, we can also establish a Noether theorem to find invariants. The theorem is stated as follows.
	\begin{theorem}[Noether theorem for good non-geometric symmetries]\label{NoG2} If $Y\in\mathfrak{X}(\mathbb R^k\times M)$ is a good  non-geometric symmetry  of order  $n$ of an exact $k$-cosymplectic Hamiltonian system $(\mathbb R^k\times M,\omega^A,\eta^A=d\theta^A,H)$, then
		\begin{align*}
			\mathcal F&=(F^1,...,F^k)\\
			&=(\zeta^1-L_Y^{n-1}i_Y\theta^1+c_{n-1}L_Y^{n-2} i_Y \theta^1+\cdots+c_1 i_Y \theta^1,...,\zeta^k-L_Y^{n-1}i_Y\theta^k+c_{n-1}L_Y^{n-2} i_Y \theta^k+\cdots+c_1 i_Y \theta^k)
		\end{align*}
		is a conserved quantity, that is, for every integrable $k$-vector field $\mathbf X=(X_1,...,X_k)\in \mathfrak{X}_H^k(\mathbb R^k\times M),$ we have that $\sum_{A=1}^kL_{X_A}F^A=0$ (on $U_p$).	
	\end{theorem}
	\begin{proof}
		If $\mathbf X=(X_1,...,X_k)\in \mathfrak{X}_H^k(\mathbb R^k\times M),$ taking \eqref{LYomm} into account we have
		$$\sum_{A=1}^kL_{X_A}F^A=\sum_{A=1}^ki_{X_A}dF^A=\sum_{A=1}^ki_{X_A}(	L_Y^{n-1}i_Y\omega^A-c_{n-1}L_Y^{n-2}i_Y\omega^A-\cdots-c_1i_Y\omega^A).$$
		Then, if $n=2,$  we have 
		\begin{align*}
			\sum_{A=1}^kL_{X_A}F^A&=\sum_{A=1}^ki_{X_A}(L_Yi_Y\omega^A-c_1i_Y\omega^A)=\sum_{A=1}^k(L_Yi_{X_A}-i_{[Y,X_A]}-c_1i_{X_A})i_Y\omega^A\\
			&=\sum_{A=1}^k(-L_Yi_Yi_{X_A}+i_Yi_{[Y,X_A]}+c_1i_Yi_{X_A})\omega^A\\
			&=-L_Yi_Y(dH-\sum_{A=1}^k(R_AH)\eta^A)+c_1i_Y(dH-\sum_{A=1}^k(R_AH)\eta^A)\\
			&=-L^2_YH+c_1L_YH=0.
		\end{align*}
		since $Y$ is a good non-geometric symmetry, $L_YH=0$ and by Definition \ref{De4}, $[Y,\mathbf X]\in Ker\omega^\sharp\cap\eta^\sharp,$ and hence $\sum_{A=1}^ki_{[Y,X_A]}\omega^A=0.$ 
		
		If $n=3,$ by an analogous reasoning, we obtain
		\begin{align*}
			\sum_{A=1}^k L_{X_A} F^A &= \sum_{A=1}^k i_{X_A} (L_Y L_Y i_Y \omega^A-c_2L_Yi_Y\omega^A-c_1i_Y\omega^A) \\
			&= \sum_{A=1}^k\left( (L_Y i_{X_A} - i_{[Y,X_A]}) L_Y -c_2(L_Yi_{X_A}-i_{[Y,X_A]})-c_1i_{X_A}\right)i_{Y}\omega^A \\
			&= \sum_{A=1}^k \left( L_Y^2 i_{X_A}  - 2L_Y i_{[Y,X_A]} + i_{[Y,[Y,X_A]]} -c_2L_Yi_{X_A}+c_2i_{[Y,X_A]}-c_1i_{X_A} \right) i_Y \omega^A \\
			&= \sum_{A=1}^k (L_Y^{2} i_{X_A}-c_2L_Yi_{X_A}-c_1i_{X_A}) i_Y \omega^A \\
			&= \sum_{A=1}^k (-L_Y^{2} i_Y+c_2L_Yi_Y+c_1i_Y) i_{X_A} \omega^A \\
			&= (-L_Y^{2} i_Y+c_2L_Yi_Y+c_1i_Y) \left( dH - \sum_{A=1}^k (R_A H) \eta^A \right) \\
			&= -L_Y^3 H+c_2L_Y^2H+c_1L_YH = 0.
		\end{align*}
		
		For $n>3,$ we arrive at the same result by repeating the above procedure $n-2$ times. In fact,
		\begin{align*}
			\sum_{A=1}^k L_{X_A} F^A &= \sum_{A=1}^k i_{X_A}  (	L_Y^{n-1}i_Y\omega^A-c_{n-1}L_Y^{n-2}i_Y\omega^A-\cdots-c_1i_Y\omega^A) \\
			&= \sum_{A=1}^k (L_Y i_{X_A} - i_{[Y,X_A]}) L_Y^{n-2} i_Y \omega^A-c_{n-1} (L_Y i_{X_A} - i_{[Y,X_A]}) L_Y^{n-3} i_Y \omega^A-\cdots-c_1i_{X_A}i_Y\omega^A \\
			&= \sum_{A=1}^k \left( L_Y^{n-1} i_{X_A}-c_{n-1}L_Y^{n-2}i_{X_A}-\cdots-c_1i_{X_A} + \tilde c_1 L_Y^{n-2} i_{[Y,X_A]} + \cdots + \tilde c_{n-1} i_{L_Y^{n-1} X_A} \right) i_Y \omega^A \\
			&= \sum_{A=1}^k \left( L_Y^{n-1} i_{X_A}-c_{n-1}L_Y^{n-2}i_{X_A}-\cdots-c_1i_{X_A} \right)  i_Y \omega^A  \\
			&= \sum_{A=1}^k \left( -L_Y^{n-1} i_{Y}+c_{n-1}L_Y^{n-2}i_Y+\cdots+c_1i_{Y} \right)  i_{X_A} \omega^A\\
			&= \left( -L_Y^{n-1} i_{Y}+c_{n-1}L_Y^{n-2}i_Y+\cdots+c_1i_{Y} \right) \left( dH - \sum_{A=1}^k (R_A H) \eta^A \right) \\
			&= -L_Y^n H +c_{n-1}L_Y^{n-1}H+\cdots+c_1L_YH= 0,
		\end{align*}
		where $\tilde c_1,...,\tilde c_{n-1}$ are  constants.
		Thus, taking into account Remark \ref{Re4}, we have proved that $\mathcal F=(F^1,...,F^k)$ is a conservation law.
	\end{proof}
	\section{Similar Results for $k$-Symplectic Manifold}
	In this section, we point out that the conclusions from the previous section can be extended to $k$-symplectic manifolds. We will not provide detailed proofs for the conclusions obtained in this section, as the proof process is similar to that of the conclusions in the previous section.
	
	Let us begin by briefly recalling the definition of Hamiltonian systems on $k$-symplectic manifolds.
	
	Consider an  \( k \)-symplectic manifold \((M,  \omega^A, V)\), and let \( H \in C^\infty(M) \) be a Hamiltonian function. The couple \((M,\omega^A, H)\) is called a \( k \)-symplectic Hamiltonian system.
	We denote by \( X_H^k(M) \) the set of (local) \( k \)-vector fields \( X = (X_1, \ldots, X_k) \) on \(  M \) which are solutions to the equations
	\begin{align}\label{HKS}
		\sum_{A=1}^k i(X_A)\omega^A = dH.	
	\end{align}
	
	Furthermore, for a map \( \psi: I \subset \mathbb{R}^k \to M \), the Hamilton-de Donder-Weyl equations for this system are
	\begin{align}
		\sum_{A=1}^k i_{ \left( \psi_*(t) \left( \frac{\partial}{\partial t^A} \Big|_t \right) \right) }(\omega^A \circ \psi) = dH \circ \psi ,
	\end{align}  	 
	Similar to the case of $k$-cosymplectic manifold, we can also define the vector bundle morphism,
	\begin{align*}
		\omega^\sharp:\;\quad T_k^1 M&\rightarrow T^*M	\\
		(X_1,...,X_k)&\rightarrow\sum_{A=1}^ki_{X_A}\omega^A.
	\end{align*}
	
	In  \cite{Roman2}, the authors proved the following Proposition.
	\begin{proposition} \cite{Roman2}\label{PRLL} 	Let $(M,\omega^A,H)$ be a $k$-symplectic Hamiltonian system.
		The map $\mathcal F=(F^1,...,F^k):\mathbb R^k\times M\rightarrow \mathbb R^k$ defines a conservation law if and only if, for every integrable $k$-vector field $\mathbf X=(X_1,...,X_k)$ which is a solution to equation \eqref{HKS}, we have
		\begin{align}
			\sum_{A=1}^kL_{X_A}F^A=0.
		\end{align}
	\end{proposition}
	Analogously to the generalized infinitesimal symmetry introduced for a \(k\)-cosymplectic Hamiltonian system in Definition \ref{De4}, here we introduce the generalized infinitesimal symmetry for a \(k\)-symplectic Hamiltonian system.
	\begin{definition}
		Let $( M,\omega^A,H)$ be a $k$-symplectic Hamiltonian system. A vector field $Y\in\mathfrak{X}( M)$ is said to be a generalized infinitesimal  symmetry if $L_Y\mathfrak{X}_H^k( M)\subset $ ker$\omega^\sharp$, i.e., for every $\mathbf X\in \mathfrak{X}_H^k( M),[Y,\mathbf X]\in $ker$\omega^\sharp$.
	\end{definition}
	By employing the generalized infinitesimal symmetry, we can further introduce the notion of a general infinitesimal Cartan or Noether symmetry of order $n$.
	\begin{definition}\label{DCar6}
		Let $( M,\omega^A,H)$ be a $k$-symplectic Hamiltonian system. A vector field $Y\in\mathfrak{X}( M)$ is said to be a general infinitesimal Cartan or Noether symmetry of order $n$ if:
		
		1. $Y$ is a generalized infinitesimal symmetry;
		
		2. \( L^{n}_Y\omega^A:=\overbrace{L_Y \cdots L_Y}^{n}\omega^A= 0 \), but \( L^m_Y \omega^A \neq 0 \), for \( m < n \);
		
		3. $ L_Y^nH=0.$ 
	\end{definition} 
	\begin{proposition}\label{Pro6}
		Let $Y\in\mathfrak{X}(M)$ be a general infinitesimal $k$-symplectic Cartan symmetry of order $n$ of an exact $k$-symplectic Hamiltonian system $( M,\omega^A=d\theta^A,H)$. Then, for every $p\in M,$ there exists an open neighborhood $U_p\ni p,$ such that:
		
		1. There exist functions $F^A\in C^\infty(U_p),$ unique up to additive constants, satisfying
		\begin{align}
			L_Y^{n-1}(i_Y\omega^A)=dF^A\quad \text{on } U_p.
		\end{align}
		
		2. There exist functions $\zeta^A\in C^\infty(U_p)$ such that $L_Y^n\theta^A=d\zeta^A$ on $U_p,$ and consequently
		\begin{align}
			\zeta^A=F^A+L_Y^{n-1}i_Y\theta^A \quad (\text{up to an additive constant on } U_p).
		\end{align}
	\end{proposition}
	\begin{proof}
		The proof parallels that of Proposition~\ref{Pro4}.
	\end{proof}
	As we said in the previous section, the concept of higher-order infinitesimal Cartan symmetries for $k$-symplectic manifolds was introduced in \cite{Roman2}. In  the following, by using higher-order general infinitesimal Cartan symmetries,  we  can get a similar result. 
	\begin{theorem}[Noether theorem for general infinitesimal Cartan symmetries] If $Y\in\mathfrak{X}(M)$ is a general infinitesimal Cartan symmetry of order $n$ of an exact $k$-symplectic Hamiltonian system $(M,\omega^A=d\theta^A,H)$, then
		$$\mathcal F=(F^1,...,F^k)=(\zeta^1-L_Y^{n-1}i_Y\theta^1,...,\zeta^k-L_Y^{n-1}i_Y\theta^k)$$ 
		is a conserved quantity, that is, for every integrable $k$-vector field $\mathbf X=(X_1,...,X_k)\in \mathfrak{X}_H^k( M),$ we have that $\sum_{A=1}^kL_{X_A}F^A=0$ (on $U_p$).	
	\end{theorem}
	\begin{proof}
		The proof follows along the same lines as the proof of  Theorem \ref{NoG}.
	\end{proof}
	Next, we  introduce non-geometric  symmetries for Hamiltonian systems on $k$-symplectic manifold.
	\begin{definition}
		Let $( M,\omega^A=d\theta^A,H)$ be an exact $k$-symplectic Hamiltonian system. A vector field $Y\in\mathfrak{X}(M)$ is said to be a non-geometric symmetry  of order  $n$ if:
		
\begin{enumerate}
		\item $Y$ is a generalized infinitesimal symmetry;
		
		\item $n$ is the first number satisfying
		\[
		L^{n}_Y\omega^A:=\overbrace{L_Y \cdots L_Y}^{n}\omega^A
		=c_0\omega^A+ c_1L_Y\omega^A+\cdots+c_{n-1}L_Y^{n-1}\omega^A,
		\]
		where $c_i\in \mathbb{R}$, $i=0,\dots,n-1$, and
		\[
		L_YH=0.
		\]
	\end{enumerate}
		
		Specially, if $c_0=0,$ we say $Y$ is a good non-geometric symmetry of order  $n$ .
	\end{definition} 
	Similar to the conclusion of Proposition \ref{Pro6}, for non-geometric symmetries of order $n$, we can obtain the conclusion of the following proposition.
	\begin{proposition}
		Let $Y\in\mathfrak{X}(M)$ be a non-geometric symmetry  of order  $n$ of an exact $k$-symplectic Hamiltonian system $(\mathbb R^k\times M,\omega^A=d\theta^A,H)$. Then, for every $p\in \mathbb R^k\times M,$ there exists an open neighborhood $U_p\ni p,$ such that:
		
		1. There exist functions $F^A\in C^\infty(U_p),$ unique up to additive constants, satisfying
		\begin{align}
			L_Y^{n-1}i_Y\omega^A-c_{n-1}L_Y^{n-2}i_Y\omega^A-\cdots-c_1i_Y\omega^A-c_0\theta^A=dF^A\quad \text{on } U_p.
		\end{align}
		
		2. There exist functions $\zeta^A\in C^\infty(U_p)$ such that $L_Y^n\theta^A=d\zeta^A$ on $U_p,$ and consequently
		\begin{align}
			\zeta^A=F^A+L_Y^{n-1}  i_Y \theta^A-c_{n-1}L_Y^{n-2} i_Y \theta^A-\cdots-c_1 i_Y \theta^A \quad (\text{up to an additive constant on } U_p).
		\end{align}
	\end{proposition}
	\begin{proof}
		The proof parallels that of Proposition~\ref{Propo4.2}.
	\end{proof}
	For good non-geometric symmetries, we can also establish a Noether theorem to find invariants. The theorem is stated as follows.
	\begin{theorem}[Noether theorem for good non-geometric symmetries] If $Y\in\mathfrak{X}(M)$ is a good  non-geometric symmetry  of order  $n$ of an exact $k$-symplectic Hamiltonian system $(M,\omega^A=d\theta^A,H)$, then
		\begin{align*}
			\mathcal F&=(F^1,...,F^k)\\
			&=(\zeta^1-L_Y^{n-1}i_Y\theta^1+c_{n-1}L_Y^{n-2} i_Y \theta^1+\cdots+c_1 i_Y \theta^1,...,\zeta^k-L_Y^{n-1}i_Y\theta^k+c_{n-1}L_Y^{n-2} i_Y \theta^k+\cdots+c_1 i_Y \theta^k)
		\end{align*}
		is a conserved quantity, that is, for every integrable $k$-vector field $\mathbf X=(X_1,...,X_k)\in \mathfrak{X}_H^k(M),$ we have that $\sum_{A=1}^kL_{X_A}F^A=0$ (on $U_p$).	
	\end{theorem}
	\begin{proof}
		The proof follows along the same lines as the proof of  Theorem \ref{NoG2}.
	\end{proof}
	\section{Conclusion}
	
	In this paper, we have systematically investigated higher-order Cartan symmetries and non-geometric symmetries within the framework of $k$-cosymplectic Hamiltonian field theory, and established corresponding Noether-type theorems that yield conservation laws. For the case k=1, the formalism exactly reproduces the results for higher-order symmetries known in classical non-autonomous cosymplectic mechanics.
	
	Our main contributions are as follows. First, by using the notion of \emph{generalized infinitesimal symmetry} for $k$-cosymplectic Hamiltonian systems,  we defined \emph{general infinitesimal Cartan (Noether) symmetries of order $n$} and \emph{good non-geometric symmetries of order $n$}, both of which generalize previously known symmetry concepts. Second, we proved two Noether-type theorems (Theorems~\ref{NoG} and~\ref{NoG2}) for these symmetry classes, providing explicit conserved quantities for exact $k$-cosymplectic Hamiltonian systems. These results extend the earlier work of Marrero et al.~\cite{Marrero} on standard Cartan symmetries and parallel the higher-order results of Rom\'an-Roy et al.~\cite{Roman2} from the $k$-symplectic setting to the $k$-cosymplectic setting.
	
	It is worth emphasizing the key differences from the existing literature. Compared with~\cite{Roman2}, which thoroughly addressed higher-order Noether symmetries for $k$-symplectic Hamiltonian systems, our work introduces two main modifications in the $k$-cosymplectic context: (i)~the standard condition $L_Y H = 0$ is relaxed to $L_Y^n H = 0$ for general infinitesimal Cartan symmetry definition, and (ii)~additional terms involving $c_i L_Y^i \omega^A$ appear in the non-geometric symmetry definition. These generalizations are not merely algebraic but lead to new classes of symmetries that were not covered by the framework of~\cite{Roman2}. In addition, the inclusion of the Reeb vector fields and the condition $i_Y \eta^A = 0$ are essential features distinguishing the $k$-cosymplectic case from the $k$-symplectic one.
	
	Furthermore, we extended all these results in a parallel manner to $k$-symplectic Hamiltonian systems (Section~5), showing that our framework is flexible enough to cover both geometric settings. This demonstrates the unifying nature of our approach to higher-order and non-geometric symmetries in classical field theories.
	
	Finally, we provided concrete examples  to illustrate the existence and applicability of the proposed symmetries, confirming that these new classes of symmetries do arise naturally and yield nontrivial conserved quantities.
	
	Several directions remain open for future work. It would be interesting to explore the physical interpretation of the generalized symmetries introduced here, particularly in the context of field theories with non-standard Lagrangians or Hamiltonians. Moreover, extending these results to singular (constrained) systems or to the Lagrangian side of the $k$-cosymplectic formalism would be a natural next step. The possible connections with integrability and superintegrability of classical field theories also deserve further investigation.
    
	\section*{Acknowledgment}
	The research of X. Zhao is supported by
	NSFC (Grant No. 12401234). M. de León acknowledges financial support from the Spanish Ministry of Science, Innovation and Universities under grants PID2022-137909NB-C21 and the Severo Ochoa Program for Centers of Excellence in R\&D (CEX2023-001347-S). 
    
    %The author would like to sincerely thank Professor Yong Li
	%for many useful suggestions and help. %The author expresses his deep gratitude to anonymous referees for their valuable comments which have improved the paper.
	$\\$
	
	\noindent$\mathbf{Conflict\;of\;interest\;statement.}$ On behalf of all authors, the corresponding author states that there is no conflict of interest.
	
	$\\$
	\noindent$\mathbf{Data\;availability.}$ Data sharing is not applicable to this article as no new data were created or analyzed in this study.

    $\\$
    \noindent $\mathbf{AI \; disclosure.}$ AI was used to assist with the review of the existing literature and to improve the language and clarity of the manuscript.

\end{document}